\documentclass[11pt]{elsarticle}
\usepackage{amsmath,amssymb}
\usepackage[margin=2.5cm]{geometry}
\usepackage{graphics,graphicx}
\usepackage{dcolumn,bm}
\usepackage{psfrag}
\usepackage{color}
\newcommand{\be}{\begin{equation}}
\newcommand{\ee}{\end{equation}}

\usepackage{graphics,graphicx}
\journal{Physica A}

\begin{document}
\def\be{\begin{equation}}
\def\ee{\end{equation}}

\def\bc{\begin{center}}
\def\ec{\end{center}}
\def\bea{\begin{eqnarray}}
\def\eea{\end{eqnarray}}
\newcommand{\avg}[1]{\langle{#1}\rangle}
\newcommand{\Avg}[1]{\left\langle{#1}\right\rangle}

\def\ie{\textit{i.\,e.,}}
\def\etal{\textit{et al.}}
\def\m{\vec{m}}
\def\G{\mathcal{G}}

\newcommand{\davide}[1]{{\bf\color{blue}#1}}
\newcommand{\gin}[1]{{\bf\color{green}#1}}
\newcommand{\bob}[1]{{\bf\color{red}#1}}

\title{Near universal values of social inequality indices in self-organized critical models}


\author{S. S. Manna${}^{a,}$\footnote{subhrangshu.manna@gmail.com}}
\author{Soumyajyoti Biswas${}^{b,}$\footnote{soumyajyoti.b@srmap.edu.in}}
\author{Bikas K. Chakrabarti${}^{c,a,d,}$\footnote{bikask.chakrabarti@saha.ac.in}}
\address
{
${}^a$Satyendra Nath Bose National Centre for Basic Sciences, Block-JD, Sector-III, Salt Lake, Kolkata-700106, India \\
${}^b$Department of Physics, SRM University - AP, Andhra Pradesh - 522502, India \\
${}^c$Saha Institute of Nuclear Physics, Kolkata - 700064, India \\
${}^d$Economic Research Unit, Indian Statistical Institute, Kolkata 700026, India
}

\begin{abstract}
      We have studied few social inequality measures associated with the sub-critical dynamical features (measured
in terms of the avalanche size distributions) of four 
   self-organized critical models while the corresponding systems approach their respective stationary critical 
   states. It has been observed that these inequality measures (specifically the Gini and Kolkata indices) exhibit nearly universal values
   though the models studied here are widely different, namely the Bak-Tang-Wiesenfeld sandpile, the Manna sandpile 
   and the quenched Edwards-Wilkinson interface, and the fiber bundle interface. These observations suggest that the self-organized 
   critical systems have broad similarity in terms of these inequality measures. A comparison with similar earlier observations
in the data of socio-economic systems with
unrestricted competitions suggest the emergent
inequality as a result of the possible proximity
to the self-organized critical states.
\end{abstract}

\maketitle

\section{Introduction}
\label{sec:introduction}
\begin{figure}[t]
\begin {center}
\includegraphics[width=7.0cm]{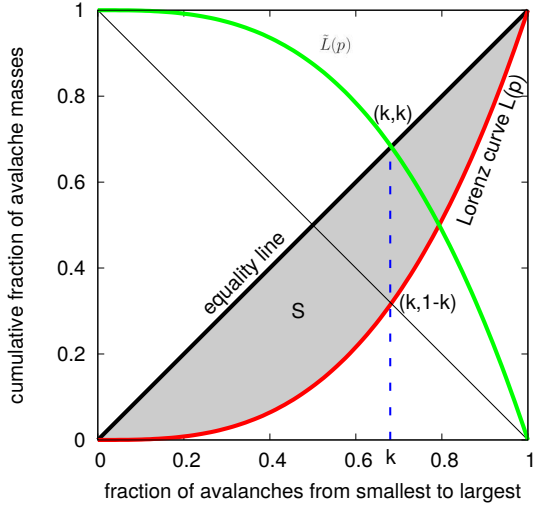}
\end {center}
\caption{The Lorenz curve $L(p)$ (shown in red) and complementary Lorenz curve $\tilde{L}(p)$ (shown in green) used to calculate  the inequality indices Gini ($g=2S$) and Kolkata (given by the fixed point $k=\tilde{L}(k)$) are shown. For the systems considered here, the horizontal axis represents the fraction 
of avalanches, when all avalanches are arranged from lowest to the highest size. The vertical axis is the fraction of the cumulative mass of these avalanches.
}
\label {first_fig}
\end{figure}

Unequal distributions of resources (for example income
or wealth) among the population are ubiquitous. Social
scientists, economists in particular, traditionally
quantify such inequalities in distributions using
some inequality indices, defined through the Lorenz
function $L(p)$ \cite{lorenz}. After ordering the population from
the poorest to the richest, the Lorenz function
$L(p)$ is given by the cumulative wealth fraction
possessed by the $p$ fraction of the population staring
from the poorest: $L(0) = 0$ and $L(1) = 1$ (see Fig. \ref{first_fig}). 
If everyone had equal share of wealth, $L(p) = p$ would be
linear (called the equality line) and the old and
still most popular inequality index, namely the Gini
($g$) index \cite{gini} is given by the ratio of the area
between the equality line and the Lorenz curve and the
entire area (1/2) below  the equality line. As such,
$g = 0$ corresponds to perfect equality and $g = 1$
corresponds to extreme inequality. Another recently
introduced  inequality index, namely the Kolkata
($k$) index \cite{ghosh14}, can be defined as the nontrivial
fixed point  of the complementary Lorenz function
$\tilde{L}(p) \equiv 1 - L(p)$: $\tilde{L}(k) = k$. It says,
$(1 - k)$ fraction of people possess $k$ fraction of
wealth ($k = 1/2$ corresponds to perfect equality
and $k = 1$ corresponds to extreme inequality). As
such, $k$ index quantifies and generalizes (see e.g.,
\cite{banerjee20}) the (more than a) century old 80-20 law ($k=0.8$) of
Pareto \cite{pareto}. Extensive analysis of social data (see
e.g., \cite{chatterjee17, ghosh21}) indicated that in extremely competitive
situations, the indices  $k$ and $g$ equals each
other in magnitude and becomes about 0.87.

In physics, starting from kinetic theory
models to that of the phase transitions, unequal
distributions of quantities like  energy, domain
sizes, avalanche sizes, etc. are widely observed. In the
renormalization group theory, we  look for the
self-similar limit (fixed point) of the generators
of such distributions, where inequalities become
extreme and one dominant (fractal) cluster emerges
and eventually brings about the phase transition.
Because of this, while the fractal dimensions
(exponents) become universal, the fixed point
itself (where only a single dominant cluster
emerges) do not become so. Social statistics, on
the other hand, look for the fixed point of the
inequality distribution, where the larger
inequalities set in, but they appear with
commensurately large frequencies (or numbers).
This some times (most likely due to 
self-organization induced by extreme competition)
leads to some universal value of  the inequality
indices (fixed points) like those of $g$ or $k$
indices discussed above.
 Indeed, a recent study
\cite{biswas21} of the $k$ index near the breaking point of the
bundle in the Fiber Bundle model, showed that $k$ attains a stable value of about 0.62 (or  inverse of the
Golden Ratio). This, however, is not a self-organized critical system 
and the terminal values of $g$ and $k$ are not equal.
 
      In this work, we consider four models where self-organized criticality (SOC) was well 
   investigated before. Specifically, we study the Bak-Tang-Wiesenfeld (BTW) \cite {btw} and Manna \cite {manna} sandpile models 
   and the driven Edwards-Wilkinson (EW) \cite{ew} and the centrally loaded fiber bundle \cite{fbm_soc} interface models. In each of these models, 
   a slow external drive brings the activity rate (toppling rate of the sand grains, or the velocity of 
   the interface) from zero to a higher value, where it saturates upon reaching the SOC state. On the 
   path to reach the SOC state, the avalanches or the cluster of activities are of unequal sizes. We 
   measure the inequality indices ($g$ and $k$) of the avalanche sizes for these models along their paths 
   towards the respective SOC states. We find that in all these four SOC models  the two indices ($g$ and $k$) become equal (about $0.87$) where the SOC sets in, 
   even though the models represent very different universality classes.

      In view of the fact that the socio-economic systems mentioned before have long been conjectured 
   to operate near the SOC state, and the inequality analysis of those systems reveal $g_c=k_c\approx 0.87$, 
   there are two main implications of our observations in the SOC models here. First, it stands as the
   quantitative evidence of social systems with unrestricted competitions operating near the SOC state. 
   Second, the time evolution of $g$ and $k$ can help in estimating the `distance' from the SOC state, 
   where such systems reach a steady state.    


\begin{figure}[t]
\begin {center}
\includegraphics[width=6.0cm]{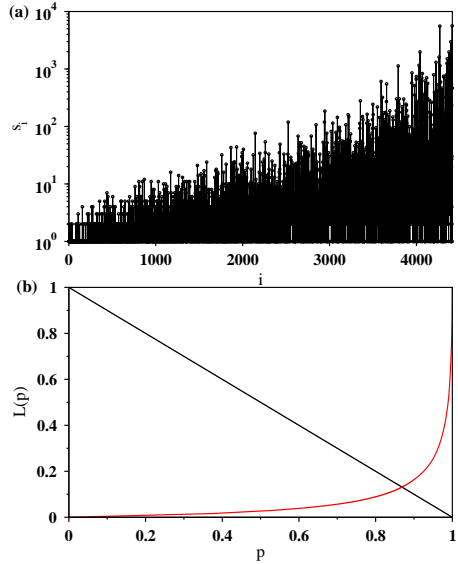}
\end {center}
\caption{A plot for the BTW sandpile model on the square lattice of size $L$ = 128 in the
subcritical phase starting from an empty lattice.
(a) The non-zero avalanche sizes $s_i$ have been plotted against their serial number $i$.
(b) The Lorenz function (cumulative avalanche size) $L(p)$ has been plotted along the 
    $y$-axis against the fraction $p$ of non-zero avalanches ordered in the sequence of 
    increasing sizes.
}
\label {FIG01}
\end{figure}
\begin{figure}[t]
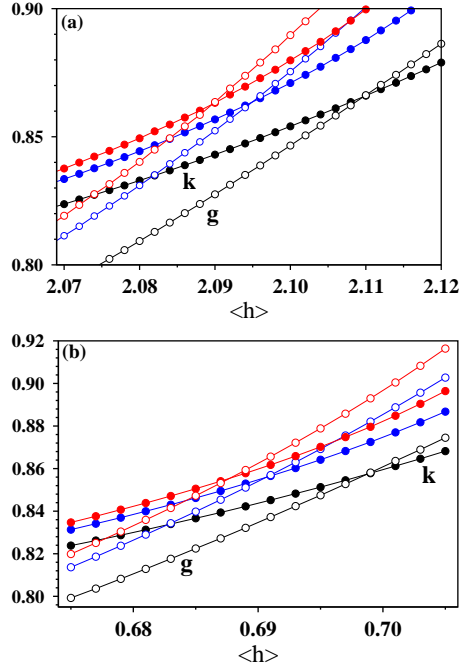

\begin {center}
\begin {tabular}{c}
\includegraphics[width=6.0cm]{BTW.eps} \\
\includegraphics[width=6.0cm]{Manna.eps} \\
\end {tabular}
\end {center}
\caption{
    (a) BTW sandpile model:
    Plot of the Kolkata index $k(\langle h \rangle, L)$ (filled circles) and the Gini coefficient 
    $g(\langle h \rangle, L)$ (empty circles) against the terminating average sandpile height 
    $\langle h \rangle$ for $L$ = 128 (black), 256 (blue) and 512 (red). For each lattice size $L$ 
    these two curves meet at a point whose $y$ coordinate is $k_c(L) = g_c(L)$ and the corresponding 
    $x$ coordinate is the average height $\langle h_c(L) \rangle$. The estimated values are:
    $k_c(L) = g_c(L) = 0.8657, 0.8634, 0.8627$ and $\langle h_c(L) \rangle = 2.110, 2.095, 2.090$
    for $L$ = 128, 256, and 512 respectively.
    (b) Manna sandpile model:
    Same quantities have been plotted for the same lattice sizes using the same symbols. The estimated values are:
    $k_c(L) = g_c(L) = 0.8573, 0.8558, 0.8555$ and $\langle h_c(L) \rangle = 0.6987, 0.6906, 0.6877$
    for $L$ = 128, 256, and 512 respectively.
}
\label {FIG02}
\end{figure}

\section {Inequality measures in the BTW and Manna sandpile models}

      In this section, we first describe the dynamics of the BTW \cite{btw} and Manna \cite{manna} 
   sandpile models. In the following sub-section we describe the method used for the estimation of
   inequality measures and report the results obtained for these models.

\subsection{Dynamics of the BTW sandpile model}

      On a square lattice of size $L \times L$ sand 
   grains are dropped one at a time at the randomly selected lattice sites with coordinates $(i,j)$. 
   Let us denote the number of sand grains at a site by $h(i,j)$. Therefore, the addition of a grain 
   implies the unit increase of the height of the sand column:
\begin {equation}
h(i,j) \rightarrow h(i,j)+1. \nonumber
\end {equation}
   A threshold value $h_c = 4$ for the height of the sand column has been pre-assigned for its stability 
   which is assumed to be the same for all sites. If the height of a sand column becomes equal to, or 
   larger than the threshold height i.e., $h(i,j) \ge h_c$, that sand column becomes unstable and it 
   topples immediately. In a toppling, the sand column looses 4 grains which are then distributed to 
   the neighboring sites uniformly:
\begin {equation}
\begin {tabular}{c}
$h(i,j) \rightarrow h(i,j)-4$ \nonumber \\
$h(i\pm1,j\pm1) \rightarrow h(i\pm1,j\pm1) + 1$. \nonumber
\end {tabular}
\end {equation}
   On receiving sand grains, some of the neighboring sites may topple again which distribute
   grains leading to further toppling. In this way, a cascade of sand column toppling takes
   place as a domino effect. The sequence of topplings is referred as an `avalanche' which 
   terminates only when none of the lattice sites remain unstable. The boundary of the lattice 
   is assumed to be open on all four sides. In the case of toppling at a boundary site, one or 
   two sand grains leave the system and never come back. A typical measure of the strength of 
   the avalanche is the total number of toppings in an avalanche and is referred as the `size' 
   $s$ of the avalanche. Usually, one starts from an empty lattice and adds sand grains one by 
   one at randomly selected sites. Therefore, the sizes of the avalanches in the early stage 
   are quite small which gradually become larger. Eventually, a stationary state is reached 
   when the outflow rate sand mass becomes equal to the inflow rate on the average. In the 
   stationary state of an infinitely large systems $L \rightarrow \infty$ the avalanche sizes 
   are expected to be of `all' length and time scales and their probability distributions are 
   characterized by the power law decays, a signature of the self-organized critical state. 
   However, for the finite size systems the avalanche sizes are limited to some $L$ dependent 
   cut-off values: $s_c(L) \sim L^{\beta}$. The average height $\langle h \rangle$ per site of 
   the sand column initially grows but eventually reaches a steady value $\langle h(L) \rangle$ 
   in the stationary state. For the infinitely large system $\langle h \rangle$ = 17/8 is an 
   analytically estimated value \cite {Levine}.

\begin{figure}[t]
\begin {center}
\includegraphics[width=6.0cm]{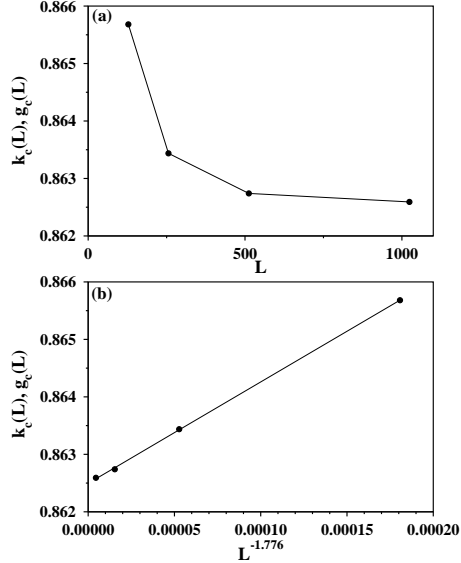}
\end {center}
\caption{BTW sandpile model:
(a) The values of the Kolkata index $k_c(L)$ and the Gini coefficient $g_c(L)$ have been plotted
    against the system size $L$ for $L$ = 128, 256, 512 and 1024. Coefficient
    values decrease on increasing the system size.
(b) The same coefficients have been extrapolated as $L^{-1/\nu}$ where the actual value of $\nu$ has
    been tuned for the best possible linear fit of the data. The plot shows that the best fit corresponds
    to $1/\nu=1.776$ which gives $k_c = g_c = 0.863$ in the limit of $L\rightarrow \infty$. 
}
\label {FIG03}
\end{figure}
\begin{figure}[t]
\begin {center}
\includegraphics[width=6.0cm]{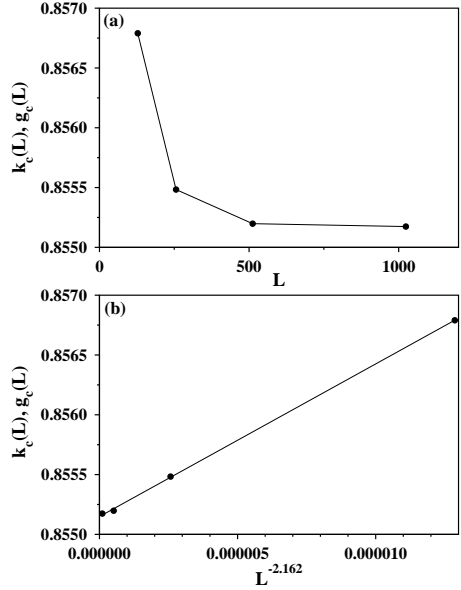}
\end {center}
\caption{Manna sandpile model:
(a) The values of the Kolkata index $k_c(L)$ and the Gini coefficient $g_c(L)$ have been plotted
    against the system size $L$ = 128, 256, 512 and 1024. Coefficient
    values decrease on increasing the system size.
(b) The same coefficients have been extrapolated as $L^{-1/\nu}$ where the actual value of $\nu$ has
    been tuned for the best possible fit of the data. The plot shows that the best fit corresponds
    to $1/\nu=2.162$ which give $k_c = g_c = 0.8554$ in the limit of $L\rightarrow \infty$. 
}
\label {FIG04}
\end{figure}
\begin{figure}[t]
\begin {center}
\includegraphics[width=6.0cm]{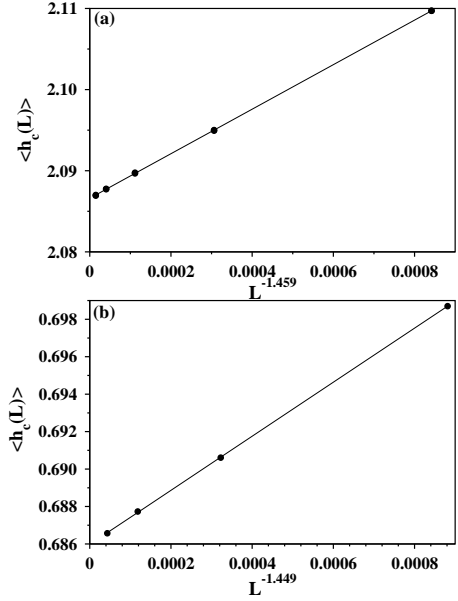}
\end {center}
\caption{
The average height $\langle h_c(L) \rangle$ of the sand column per site corresponding to the 
point where curves for $k(\langle h \rangle,L)$ and $g(\langle h \rangle,L)$ meet in Fig. 2
have been plotted for $L = 128, 256, 512, 1024$ and $2048$.
(a) BTW sandpile model: The values of $\langle h_c(L) \rangle$ have been
    plotted against $L^{-1.459}$ which is extrapolated to $h_c = 2.087$ in the limit of $L \rightarrow \infty$.
    This value is somewhat smaller than the value of the average height 17/8 per site for BTW model with $L \rightarrow \infty$ \cite {Levine}. 
(b) Manna sandpile model: The values of $\langle h_c(L) \rangle$ have been
    plotted against $L^{-1.449}$ which is extrapolated to $h_c = 0.6859$ in the limit of $L \rightarrow \infty$.
    This value is somewhat smaller than the value of the average height $\approx$ 0.717157 per site for Manna model with $L \rightarrow \infty$ \cite {MannaCal}. 
}
\label {FIG05}
\end{figure}

\begin{figure}[t]
\begin {center}
\includegraphics[width=6.0cm]{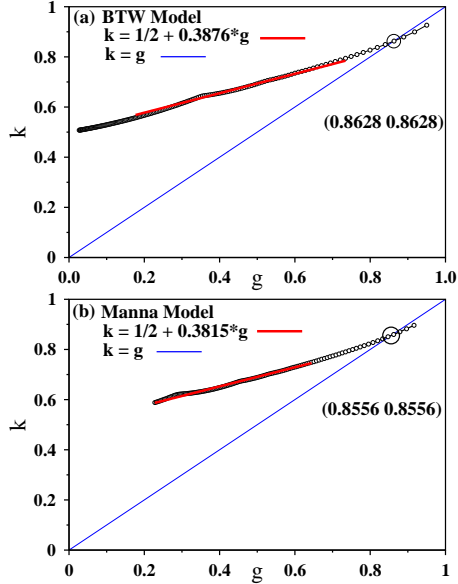}
\end {center}
\caption{
The variations of $k$ versus $g$ in the
(a) BTW sandpile model, and
(b) Manna sandpile model. The initial parts follow a straight line with slightly different slopes, as indicated in the
figures. The crossing points with the $g=k$ line are 0.8628 and 0.8556 respectively.
}
\label {g_k_sandpile}
\end{figure}

\subsection {Dynamics of the Manna sandpile model}

      This is a variant of the BTW sandpile model where the threshold height for stability
   is assumed to be $h_c = 2$. When a sand column topples, it looses only 2 grains which are
   then distributed to the four neighboring sites, each grain is transferred to one of the 
   neighboring sites, selected randomly.
\begin {equation}
\begin {tabular}{c}
$h(i,j) \rightarrow h(i,j)-2$ \nonumber \\
$h(i_n,j_n) \rightarrow h(i_n,j_n) + 1$, \nonumber
\end {tabular}
\end {equation}
   where $(i_n,j_n)$ is a randomly selected neighboring site and the second equation is executed
   twice. Using the similar open boundary condition, one finds the average height per site 
   in the stationary state is approximately 0.717157 \cite {MannaCal}. 

\subsection {Inequality analysis of the sandpile models}

      We have studied both the BTW and Manna sandpile models in the subcritical state. Starting
   from a completely empty lattice, sand grains have been added on to the lattice sequentially 
   one after another. In general, each sand grain addition results an avalanche whose size may 
   be zero or larger. The data for only the non-zero avalanche sizes have been collected till 
   the average height of the sand column per site reaches a pre-assigned value. The total number 
   $N$ of the non-zero avalanches depend on the value of the terminating average height $\langle h(L) \rangle$ 
   where we stop the simulation. In Fig. \ref {FIG01}(a) we have plotted this fluctuating sequence 
   of the avalanche sizes $s_i$ as they occur during the sand dropping process, against their 
   serial number $i$. The avalanche cluster sizes are then sorted out 
   in an ordered increasing sequence $\{s_i, i=1,N\}$. Further, we have defined a quantity $p$ which 
   is the fraction of avalanches, starting from the smallest to the largest. Simultaneously, we have
   defined the Lorenz function $L(p) = \Sigma_{i=1}^{i}s_i / \Sigma_{i=1}^{N}s_i$ \cite{lorenz} which represents 
   the fractional cumulative size and its value has been plotted against $p$ (Fig. \ref {FIG01}(b)). 
   On this plot both the axes have been normalized to unity. We also draw the $y = 1-x$ line and 
   the value of the Kolkata index $k$ is obtained by the value of $p$ coordinate of the point of 
   intersection of the $L(p)$ vs. $p$ curve (shown in red) and the $y = 1-x$ line. In addition, 
   the Gini coefficient $g$ is obtained by doubling the area between the Lorenz curve and 
   the diagonal $y = x$ line.

      The values of the Kolkata index $(k)$ and the Gini coefficient $(g)$ have been plotted for the
   BTW and Manna sandpile models in Fig. \ref {FIG02}(a) and Fig. \ref {FIG02}(b) respectively.
   The entire calculation have been done for the four different system sizes, namely, $L$ = 128, 256, 
   512 and 1024; and repeated for 25 values of the terminating average height $\langle h(L) \rangle$ 
   at an interval of 0.002. For every lattice size we have plotted $k(\langle h(L) \rangle,L)$
   and $g(\langle h(L) \rangle,L)$ against $\langle h(L) \rangle$. For every lattice size, these two 
   curves intersect at a point which has the coordinates $(\langle h(L) \rangle,k_c(L) = g_c(L))$.

      The values of the system size dependent indices are then extrapolated to obtain their asymptotic values in the limit
   of infinite system size. First, the values of $k_c(L) = g_c(L)$ have been plotted against $L$ in
   Fig. \ref {FIG03} (a) for the BTW model. In the lower panel Fig. \ref {FIG03}(b) the same 
   values have been plotted against $L^{-1/\nu}$ and the value of $\nu$ has been tuned so that the
   data fits best to a linear least square fit. We obtain $1/\nu = 1.776$ and asymptotic values of 
   the indices $k_c = g_c = 0.863$ in the limit of $L \rightarrow \infty$. A similar plot in the 
   Fig. \ref {FIG04}(a) and Fig. \ref {FIG04}(b) for the Manna sandpile yields the best fitted 
   results $1/\nu = 2.162$ and $k_c = g_c = 0.855$ in the limit of $L \rightarrow \infty$. A similar
   method of data analysis had been used to find the asymptotic value of the terminating average 
   height $\langle h(L) \rangle$. In Fig. \ref {FIG05}(a) the average heights per site of the BTW
   model had been extrapolated against $L^{-1.459}$ to obtain the best linear fit. In the limit
   of $L \rightarrow \infty$, we get $h_c=2.087$. This value is somewhat smaller than the value of 
   the average height 17/8 \cite {Levine} per site for the BTW model in the limit of $L \rightarrow \infty$.
   Similarly, in Fig. \ref {FIG05}(b) we have extrapolated the same data of the Manna sandpile and 
   on extrapolation against $L^{-1.449}$ we obtained $h_c=0.6859$ to be compared with $h_c \approx 0.717157$
   \cite {MannaCal}.

   In Figs. \ref{g_k_sandpile}(a) and  \ref{g_k_sandpile}(b), $k$ vs $g$ are plotted for the BTW and Manna
   models respectively. The initial part of the plots fit a straight line, with slopes 0.3876 and 0.3815
   in the BTW and Manna models respectively. The plots intersect the $k=g$ line at 0.8628 and 0.8556 
  for the two models (for system size $L\times L=512 \times 512$).

\begin{figure}
\begin {center}
\includegraphics[width=7.0cm]{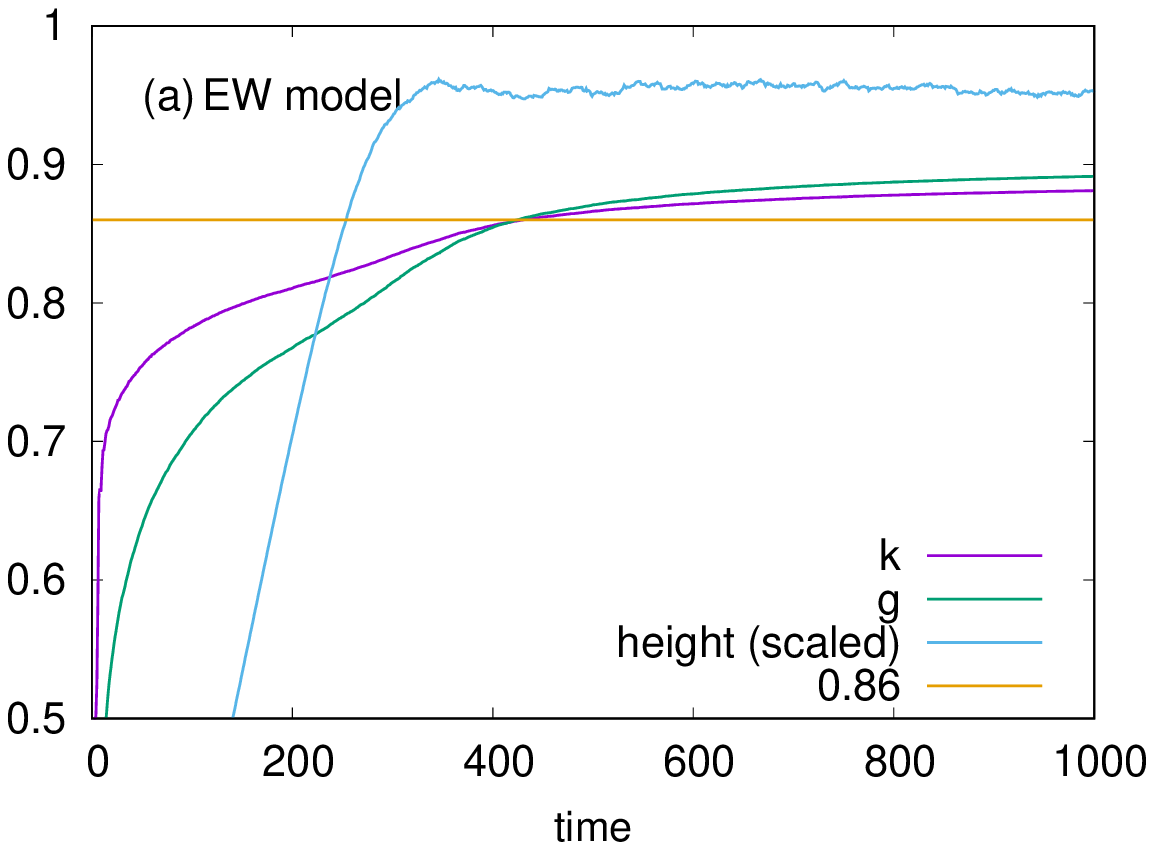}
\includegraphics[width=7.0cm]{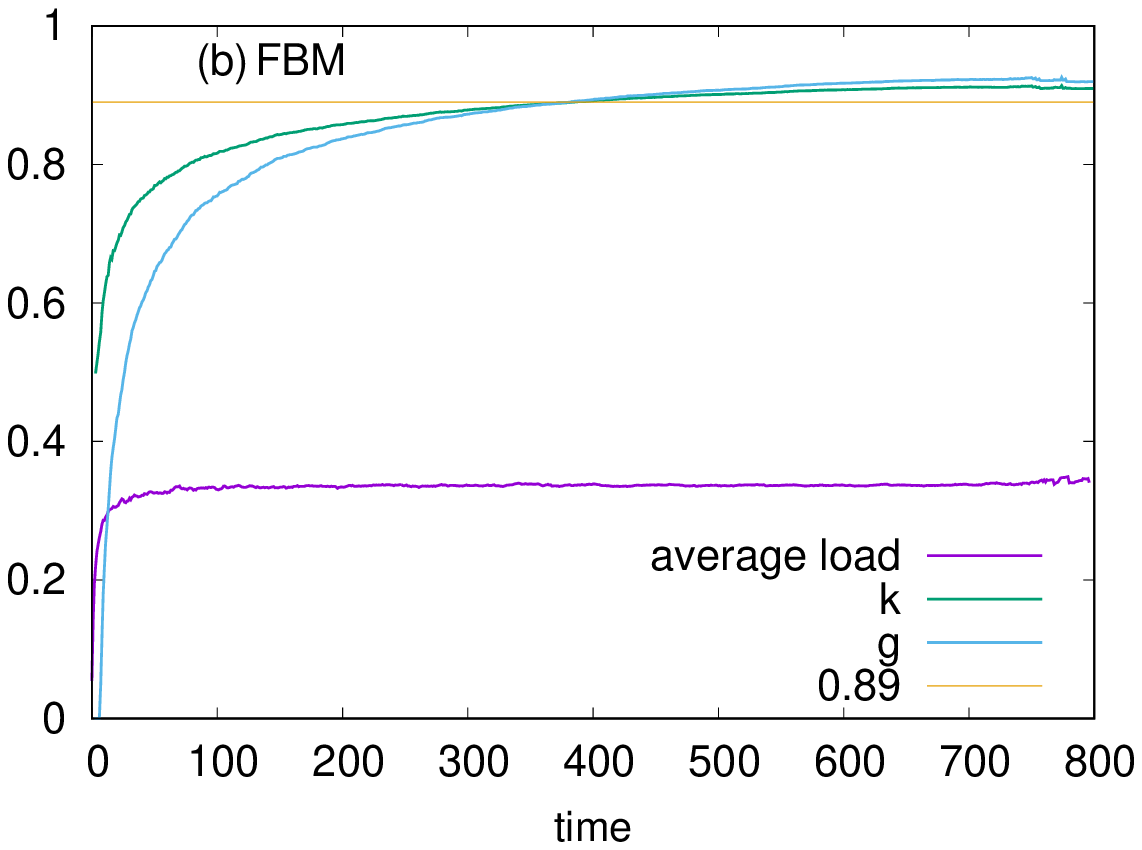}
\end {center}
\caption{The time variations of $k$ and $g$ indices in the avalanches of the EW and centrally loaded fiber bundle models approaching SOC state.
(a) The variations of the $k$ and $g$ indices are shown as functions of time for the EW model. The two indices cross
near $0.86$. This is close to, but not equal to, the time at which the height difference between the boundary points (where the drive is applied) and the middle point 
(farthest from the drive) of the EW chain reaches saturation. (b) The same is shown for the centrally loaded fiber bundle model.}
\label{k_g_time_EW_FBM}
\end{figure}

\section{Inequality measures in the boundary driven Edwards-Wilkinson and centrally loaded fiber bundle models}
Quasistatically driven interfaces through a quenched disordered medium can show self-organized critical behavior. 
Here we analyze two such models, namely the Edwards-Wilkinson mode \cite{ew} and the centrally loaded fiber bundle model \cite{fbm_soc}.
Below we first describe the dynamics of these models and then report the inequality measures in the avalanche sizes leading to the critical point.

\subsection{Dynamics of the quenched Edwards-Wilkinson model}

      The Edwards-Wilkinson (EW) model \cite{ew} in 1+1 dimension is the simplest model for driven 
   interfaces through random media. For an interface height $h(x,t)$, it reads
\begin{equation}
\frac{\partial h(x,t)}{\partial t}=\nu\nabla^2h(x,t)+\eta(x,h)+F,
\end{equation}
   where $\eta(x,h)$ is a quenched noise (pinning force) and $F$ is the driving force. The
   dynamics of the model can be represented in the discrete form as follows
   $h_i(t+1)=h_i(t)+1$ if $\tau_i>0$, where $\tau_i=h_{i+1}(t)+h_{i-1}(t)-2h_i(t)+\eta_i(h)+F$.
   The random pinning forces $\eta_i(h)$ are drawn from some probability distribution. 

While a depinning tradition can be observed for $F>F_c$, we study here the self-organized
critical dynamics of the model, which is driven only at the boundaries \cite{biswas13}.
We take a chain of length $L$ and in each stable configuration, move the $0$th and $L-1$st
points by one unit. There is n other external force anywhere else on the chain. 
The drive may cause instabilities i.e., other points may move, resulting in an avalanche. 
The total number of steps moved by all elements until the next stable configuration is reached,
is defined as the avalanche size $s$. Following some transient behavior, the model reaches
a self-organized critical state, where the height difference between the end points and the middle point 
of the chain  reaches a stationary value and fluctuates around it. The size distribution of
avalanches follow a power law.

\begin{figure}
\begin {center}
\includegraphics[width=7.0cm]{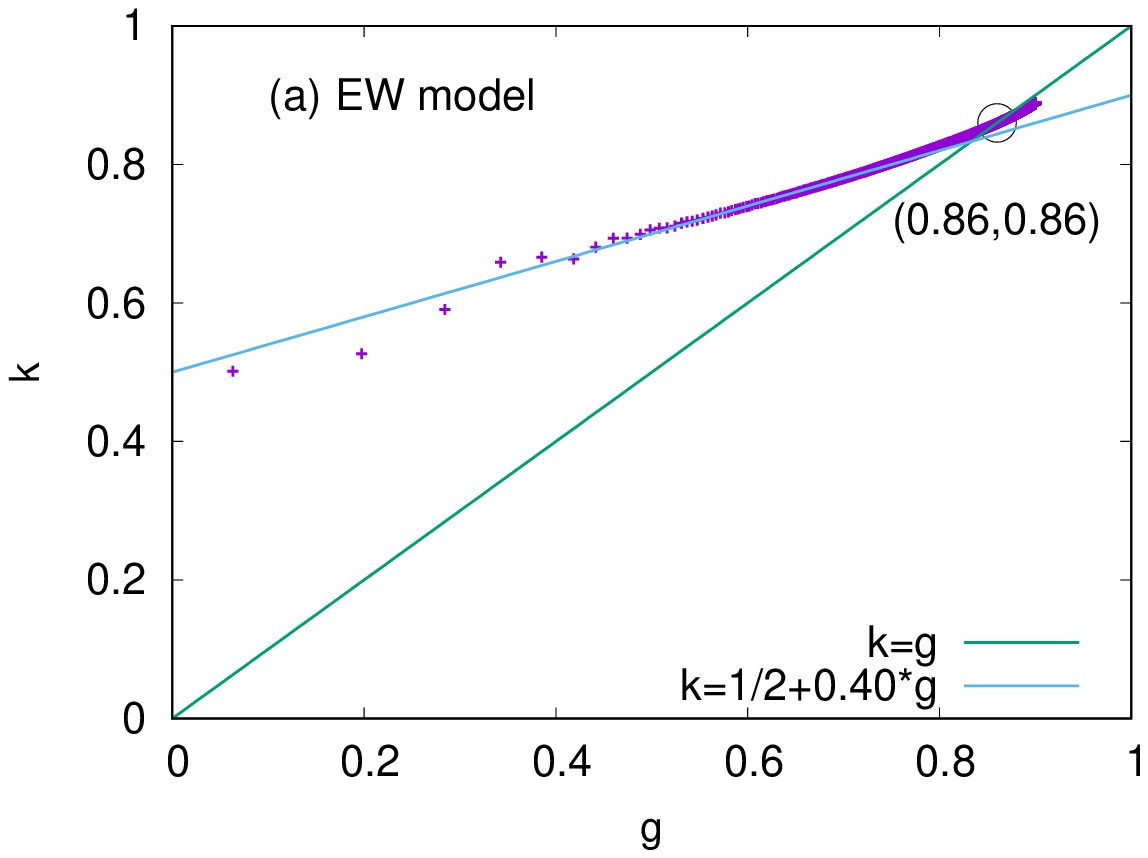}
\includegraphics[width=7.0cm]{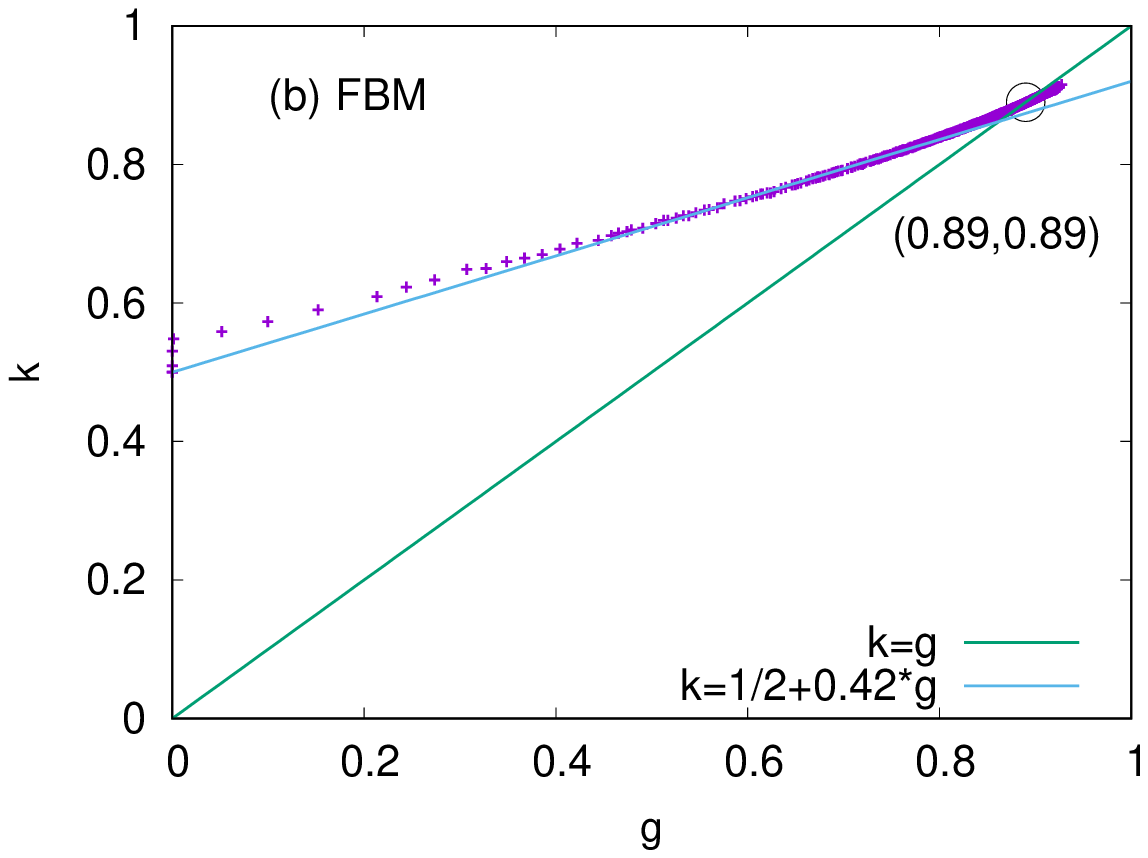}
\end {center}
\caption{The variations of $k$ versus $g$ in the EW and centrally loaded fiber bundle models.
(a) The variations of $k$ versus $g$ indices are shown for the EW model. The initial slope
of the $k$ versus $g$ plot is $0.40$. (b) The same is plotted for the centrally loaded fiber bundle model.
The slope here is $0.42$.}
\label{k_vs_g_EW_fbm}
\end{figure}
\subsection{Dynamics of the centrally loaded fiber bundle model}
The fiber bundle model \cite{fbm_rmp} is a generic model for failures in disordered materials. For almost a century, the model have been
studied by engineers and physicists in the context of catastrophic breakdown processes in driven disordered materials. 
The centrally loaded fiber bundle model \cite{fbm_soc}, however, does not lead to catastrophic breakdown, but reaches a steady state. 
The model consists of linear elastic fibers, arranged in a square grid, that are clamped between two plates. The fibers are linear elastic, until
the load on it reaches a threshold, beyond which it breaks irreversibly. While the elastic moduli of all fibers are the same, the failure thresholds
are in general different and are drawn from a distribution function. The width of the distribution function is a measure of the disorder in the system. 

 The top plate, from which the fibers are hanging, is rigid, but 
the bottom plate is very soft. Then a load is applied on a central location, whih is carried by one centrally located fiber. When the load is gradually 
increased, it crosses the failure threshold of the fiber and the fiber breaks. The load is then carried by it's four nearest neighbors. If that leads to
breaking of any one of those neighboring fibers, then the load carried by that fiber is redistributed equally between the surviving neighbors of
that broken fiber. For any given load, the centrally located damaged area will continue to increase until a point when all fibers along the damage boundary carry
 loads that are less than their respective failure thresholds. The load on all fibers along the damage boundary is then increased uniformly until the point 
when one more fiber breaks and the dynamics is restarted. The numbers of fibers breaking between two successive stable states is the size of the avalanche. 
Given that the damage boundary, hence the numbers of fibers carrying the applied load will keep on increasing, the model will never show catastrophic breakdown in
the thermodynamic limit. It turns out that the average load per fiber reaches a value $1/3$, if the threshold distribution is uniform in (0,1). This is a self-organized
critical point and the avalanche sizes show a power law distribution with exponent value $-3/2$ \cite{fbm_soc}. 

Here we look at the avalanches, starting from the breaking of the first centrally located fiber, as the system reaches the self-organized critical point and 
measure inequalities in the avalanche sizes. 

\subsection{Inequality analysis of the interface models} 


 In Fig. \ref{k_g_time_EW_FBM}(a), 
the averaged values of $g$, $k$ and the
height difference between the boundary points and the middle point scaled by a factor, are shown for
uniform distributions of the pinning forces in (0,2) in the EW model. The crossing points of $g$ and $k$ are surprisingly close to 
0.86. The same observation is found for triangular and Gaussian distributions of the pinning forces (not shown). 
In Fig. \ref{k_g_time_EW_FBM}(b), we plot the same for the centrally loaded fiber bundle model. The crossing point in this
case is slightly higher (0.89).

In Fig. \ref{k_vs_g_EW_fbm}(a) $k$ versus $g$ is plotted for the EW model. The initial part is linear, with a slope $0.40$. 
In Fig. \ref{k_vs_g_EW_fbm}(b) the same is plotted for the centrally loaded fiber bundle model. In this case, the initial linear
slope is $0.42$.

Other than the surprisingly close values of the crossing points of $g$ and $k$ in the various different models, 
the linear variation with a near-universal slope in $k$ versus $g$ plots seems to be rather crucial. 
Indeed, in the rel data for diverse socioeconomic systems, such behavior have been already observed \cite{ghosh21}.
This is particularly important in view of the fact that many socioeconomic systems with unrestricted competition 
are assumed to be in an SOC state.  
\section{Discussions and conclusions}
      In socio-economic systems with unrestricted competitions, emergence of inequality in resource 
   distributions is observed very often \cite{ghosh21,banerjee21}. Such inequalities are quantified using some indices, related 
   to the distributions of the individual resources through the Lorenz function $L(p)$. Here we have studied 
   the Gini index $g$ (given by the ratio of the area between the equality line and the Lorenz curve and the
area under the equality line), and the Kolkata index $k$ (given by the fixed point of the
complementary Lorenz function $\tilde{L}(p)\equiv 1-L(p)$), giving the magnitude  of the fraction $k$ 
   of resource possessed by the fraction $1-k$ of the population (see Fig. \ref{first_fig}). 
It was seen before that $g_c=k_c\approx 0.87$ 
   in systems with unrestricted competitions (see e.g., \cite{ghosh21,banerjee21}). 

      We have shown here, through numerical analysis, that the same behavior are shown 
   by four self-organized critical models, namely the BTW sandile \cite{btw}, Manna sandpile \cite{manna}, quenched 
EW model \cite{ew} and the centrally loaded fiber bundle model \cite{fbm_soc}, 
in their course towards the respective self-organized critical points. 
Particularly, from the time series of the avalanches in these models, we first construct the Lorenz curve (see Fig. \ref{first_fig}).
Then from its fixed point the inequality index $k$ is estimated and from the area between the equality line and the Lorenz curve the value of $g$ is estimated. The Lorenz curve is discrete here and the corresponding error
bars in $g$ and $k$ depends upon the point spacings, which in turn depend upon the system size. 
Finite size scaling 
   reveals that the crossing points of $g$ and $k$ as functions of the average sandpile height differ only 
   slightly (less than 3\%) from the respective critical heights. Specifically, for the BTW model, the crossing
point for $k$ and $g$ is at $k_c = g_c = 0.863$
(Fig. \ref{FIG02}) when  extrapolated to $L \rightarrow
\infty$ while the  average height of the BTW
sandpile at that point is 2.087  (Fig. \ref{FIG04}), which
is slightly less than the critical height \cite{Levine} of the
BTW sandpile (17/8).  For the Manna sandpile
model, the crossing point of $k$ and $g$ is at
$k_c = g_c = 0.8554$ (Fig. \ref{FIG03}) The average sandpile height at that point 
is 0.6859 (Fig. \ref{FIG04}), while the critical height is at $0.717157$ \cite{MannaCal} for the Manna model. Also, the slope of the $k$ versus $g$
line at the initial stage (prior to reaching SOC state) is 0.3876 and 0.3815 for the BTW and Manna model (Fig. \ref{g_k_sandpile}) respectively.
In the EW interface, the crossing point of $k$ versus $g$ is $k_c=g_c \simeq 0.86$ and that for the centrally loaded fiber bundle   
 model is about 0.89 (Fig. \ref{k_g_time_EW_FBM}). The initial linear part of the $k$ versus $g$ plots have slope 0.40 and 0.42 respectively for the 
EW and the centrally loaded FBM (Fig. \ref{k_vs_g_EW_fbm}). A comparison with similar observations in
socio-economic data (see e.g., \cite{banerjee21}) suggests
similar  emergent inequality as a result of
unconstrained competitions in the social dynamics. 

The underlying physical origin of this numerical observation, seen here in SOC models of distinct universality classes, and also seen elsewhere in 
socio-economic systems \cite{ghosh21,banerjee21} presumably operating in SOC points, is an important open question. While some recent progresses have 
been made in analytically (see e.g., \cite{bijin}), further extensive analysis both in models and data (e.g., see \cite{perc}) are needed.   

     In conclusions, the self-organized critical (SOC) models
we studied here show near universal characteristics in terms
of the inequality measures of their `avalanche' statistics in
their  respective approach to the SOC points. This stands
as a quantitative support  the existence  of SOC behavior
in socio-economic systems. Such observations can also help
in estimating the `distance' from the imminent disaster or SOC points for
those complex  systems for which such model studies are
not possible.
\vskip0.5cm
\noindent {\it Acknowledgement:} SSM is thankful to S. N. Bose National Centre for Basic Sciences,
Kolkata for the support through the Visiting (Honorary) Fellow position. BKC is grateful to the 
Indian National Science Academy for their support through the Senior Scientist Research Grant.

\end{document}